\documentclass[aps,prl, twocolumn,superscriptaddress,nofootinbib,notitlepage,showpacs,floatfix]{revtex4-1}

\usepackage{graphicx,graphics,epsfig,subfigure,times,bm,amssymb,amsmath,amsfonts,amsthm,mathrsfs,MnSymbol}
\usepackage[matrix,frame,arrow]{xypic}
\usepackage[pdfstartview=FitH]{hyperref}
\usepackage{pifont}
\hypersetup{
    colorlinks=true,       % false: boxed links; true: colored links
    linkcolor=red,          % color of internal links
   citecolor=magenta,        % color of links to bibliography
    filecolor=magenta,      % color of file links
    urlcolor=cyan,           % color of external links
    runcolor=cyan
}
\usepackage[pdftex]{color}
\usepackage{braket}%Dirac Notation in QM
\usepackage{enumerate}
\usepackage[normalem]{ulem}
\usepackage{pifont}% http://ctan.org/pkg/pifont
\usepackage{array}
\newcommand{\cmark}{\ding{51}}%
\newcommand{\xmark}{\ding{55}}%
\newcolumntype{M}[1]{>{\centering\arraybackslash}m{#1}}

\usepackage{subfigure}
\usepackage{overpic}

\usepackage{multirow}
\newcommand{\be}{\begin{equation}}
\newcommand{\ee}{\end{equation}}
\newcommand{\ba}{\begin{eqnarray}}
\newcommand{\ea}{\end{eqnarray}}

\newcommand{\ignore}[1]{}

\begin{document}

\title{Entanglement Complexity in Quantum Many-Body Dynamics,
  Thermalization and Localization}

\author{Zhi-Cheng Yang}
 \affiliation{Physics Department, Boston University, Boston,
   Massachusetts 02215, USA}

\author{Alioscia Hamma}
\affiliation{Physics Department, University of Massachusetts, Boston,
  Massachusetts 02125, USA}
 
\author{Salvatore M. Giampaolo}
\affiliation{International Institute of Physics, Universidade Federal
  do Rio Grande do Norte, 59078-400 Natal-RN, Brazil}

\author{Eduardo R. Mucciolo}
\affiliation{Department of Physics, University of Central Florida,
  Orlando, Florida 32816, USA}

\author{Claudio Chamon} 
\affiliation{Physics Department, Boston University, Boston,
  Massachusetts 02215, USA}

\begin{abstract}
Entanglement is usually quantified by von Neumann entropy, but its
properties are much more complex than what can be expressed with a
single number. We show that the three distinct dynamical phases known
as thermalization, Anderson localization, and many-body localization
are marked by different patterns of the spectrum of the reduced
density matrix for a state evolved after a quantum quench. While the
entanglement spectrum displays Poisson statistics for the case of
Anderson localization, it displays universal Wigner-Dyson statistics
for both the cases of many-body localization and thermalization,
albeit the universal distribution is asymptotically reached within
very different time scales in these two cases. We further show that
the complexity of entanglement, revealed by the possibility of
disentangling the state through a Metropolis-like algorithm, is
signaled by whether the entanglement spectrum level spacing is Poisson
or Wigner-Dyson distributed.
\end{abstract}

\maketitle

{\em Introduction.---} Entanglement is usually quantified by a number,
the entanglement entropy, defined as the von Neumann entropy of the
reduced density matrix $\rho_A$ of a subsystem, and it is a key
concept in many different physical settings, from novel phases of
quantum matter~\cite{ref48,ref49,ref50,ref51} to
cosmology~\cite{bh,susskind1}. However, there is a lot more
information in the entanglement spectrum of $\rho_A$, namely the full
set of its eigenvalues (or its logarithms)~\cite{haldane-li}.
Recently, a measurement protocol to access the entanglement spectrum
of many-body states using cold atoms has been
proposed~\cite{measure}. The main goal of this letter is to explore
the relationship between entanglement spectrum and dynamical behavior
of a quantum many-body system.

In Refs.~\cite{espectrum1,espectrum2} it was shown that the
entanglement of a state generated by a quantum circuit can be simple
or complex, in the sense that the state either can or cannot be
disentangled by an {\em entanglement cooling} algorithm that resembles
the Metropolis algorithm for finding the ground state of a
Hamiltonian. The success or failure of the disentangling procedure is
signaled by the so called entanglement spectrum statistics
(ESS)~\cite{espectrum1,espectrum2}, namely the distribution of the
spacings between consecutive eigenvalues of $\rho_A$. When such a
distribution is Wigner-Dyson (WD), the cooling algorithm fails. This
situation occurs when the gates in the circuit are sufficient for
universal computing, either classical or quantum. On the other hand,
for circuits that are not capable of universal computing, the states
can be disentangled and they feature a (semi-)Poisson ESS.

In this letter, we focus on systems whose dynamics is controlled by a
time-independent quantum many-body Hamiltonian, as opposed to a random
circuit. We study the entanglement complexity revealed by the ESS of
the time-evolved state for Hamiltonians whose eigenstates yield one of
three behaviors: 1) eigenstate thermalization
(ETH)~\cite{gemmer,popescu,lloyd,rigol,reimann,goldstein}, 2) Anderson
localization (AL), or 3) many-body localization
(MBL)~\cite{mbl1,mbl2,mbl3}. We find that the time-evolved states
under Hamiltonians that feature AL follow a Poisson ESS, and that they
can be disentangled by applying the entanglement cooling algorithm
which uses only the unitaries generated from one-and two-body terms in
the Hamiltonian. On the other hand, the time-evolved states under
Hamiltonians that satisfy ETH follow a WD distribution, and the
entanglement cooling algorithm fails. Remarkably, the dynamics
generated by MBL Hamiltonians results in ESS approaching
asymptotically in time a WD distribution, the same distribution that
time-evolved states with ETH Hamiltonians reach in shorter times. We
find that the rate of such approach to WD scales with the inverse of
the logarithm of time. We further find that the state generated by MBL
Hamiltonians cannot be disentangled using a cooling algorithm.

{\em Quantum Quench of the Heisenberg spin chain.---} We shall focus
on a quantum state that is time-evolved after a {\em quantum quench},
namely, a sudden switch of the Hamiltonian so as to throw the initial
state away from equilibrium. We consider the XXZ spin-1/2 chain of $L$
sites with open boundary conditions,
\be\label{model}
 H^{} = J\sum_{i=1}^{L-1} (\sigma_i^x \sigma_{i+1}^x + \sigma_i^y
\sigma_{i+1}^y +\Delta  \sigma_i^z \sigma_{i+1}^z +
z_i \sigma_i^z +x_i \sigma_i^x )
\;.
\ee
We consider three distinct regimes of parameters: (i) In the absence
of a transverse field and interaction ($\Delta=x_i=0, z_i \neq 0$),
the Hamiltonian in Eq.~(\ref{model}) maps onto free fermions via a
Jordan-Wigner transformation~\cite{XX, barouch}. The complexity of the
problem is reduced from that of diagonalizing a $2^L\times 2^L$ matrix
to that of diagonalizing a $L\times L$ matrix. In the limit case of no
disorder, $z_i= \mbox{const}$, the system is completely integrable
while in the presence of disorder it shows AL~\cite{xxanderson}. In
the case of AL, the Hamiltonian is noninteracting in the basis of
local conserved quantities. The presence of constants of motion
prevents the system from thermalizing. (ii) In the presence of
interactions and weakly disordered external fields ($z_i\in[-1,1]$ and
$\Delta=0.5$), the Hamiltonian in Eq.~(\ref{model}) is nonintegrable
and thermalizes. Its eigenstates obey ETH. (iii) Finally, in the
presence of interactions and strong disorder ($z_i\in[-10,10]$ and
$\Delta=0.5$), the system features MBL: Even the high-energy
eigenstates of such a system are weakly entangled, obey an area law
and thus do not follow ETH~\cite{eth1,eth2,rigol}. The dynamical
behavior of the MBL phase is also apparent in the fact that during the
evolution, the entanglement grows only logarithmically in
time~\cite{marko,Moore,abaninlog}.

The quantum evolution is studied as follows. We consider the state
$|\Psi(t)\rangle =\exp(-\dot{\imath} H t)|\Psi_0\rangle$, where
$\ket{\Psi_0} = \otimes_j \ket{\psi}_j$ is a random factorized state.
By quenching to different values of $\{ x_i, z_i,\Delta\}$, we can
obtain all possible dynamics we want to study. The marginal state
$\rho_A(t)$ corresponds to the reduced density matrix of one half of
the total chain. The set of eigenvalues of $\rho_A$ are then denoted
by $\{p_i\}_{i=1}^{2^{L/2}}$ and ordered in decreasing order. At the
same time, we also consider the eigenenergies $\{E_j\}_{j=1}^{2^L}$ of
the full Hamiltonian.

{\em Entanglement spectrum statistics.---} At $t=0$, the state
contains initially no entanglement and gets entangled only through the
dynamics. After a time $t_0=1000$ in units of $1/J$, we study the ESS
of the spectrum $\{p_i\}_{i=1}^{2^{L/2}}$\cite{espectrum1,espectrum2},
here obtained from the distribution $P(r) = R^{-1} \sum_{i=1}^R
\langle\delta(r-r_i)\rangle$ of the ratios of consecutive spacings,
$r_i =(p_{i-1}- p_{i})/(p_{i}- p_{i+1})$. In an analogous fashion, we
obtain the statistics of ratios of the energy spectrum
$\{E_j\}_{j=1}^{2^L}$ and compare it to the ESS. Our results are
summarized in Table~\ref{table1}.

%%%%%%%%%%%%%%%%%%%%%%%%%%%%%%%%%%%%%%%%%%%%%%%%%%%%%%%%%%%%%%%%%%%%%%%
\begin{table}[t]
\centering
\begin{tabular}{ |M{3.2cm}||M{1.2cm}|M{1.2cm}|M{1.2cm}|  }
 \hline
 &\multicolumn{3}{c|}{Dynamical phases } \\
 \hline
Features& AL &ETH & MBL\\
 \hline 
 \hline
 Entanglement spectrum   & Poisson   & WD &   WD \\
 \hline
 Energy spectrum &   Poisson  &  \begin{tabular}{c} Poisson \\ {\it or} WD\end{tabular}   &Poisson\\
 \hline
 Entanglement cooling   &\cmark & \xmark&  \xmark \\
 \hline
\end{tabular}
\caption{Summary of the main results presented in the paper. The ESS
  of Hamiltonians featuring AL shows a Poisson distribution, while for
  both ETH and MBL Hamiltonians it displays a WD distribution. In
  particular, the deviation from the WD distribution in the MBL case
  decays as $1/\log (t)$. The energy level spacing statistics yields a
  Poisson distribution for both AL and MBL, while for ETH case it can
  be either Poisson (in the presence of additional conserved
  quantities) or WD (with no conserved quantities). Finally, the
  states generated by AL Hamiltonians can be disentangled using an
  entanglement cooling algorithm, while the states generated by ETH
  and MBL Hamiltonians cannot.}
\label{table1}
\end{table}
%%%%%%%%%%%%%%%%%%%%%%%%%%%%%%%%%%%%%%%%%%%%%%%%%%%%%%%%%%%%%%%%%%%%%%%

\begin{figure*}[!ht]
\centering
\scalebox{.45}{\includegraphics[width=2.25\textwidth]{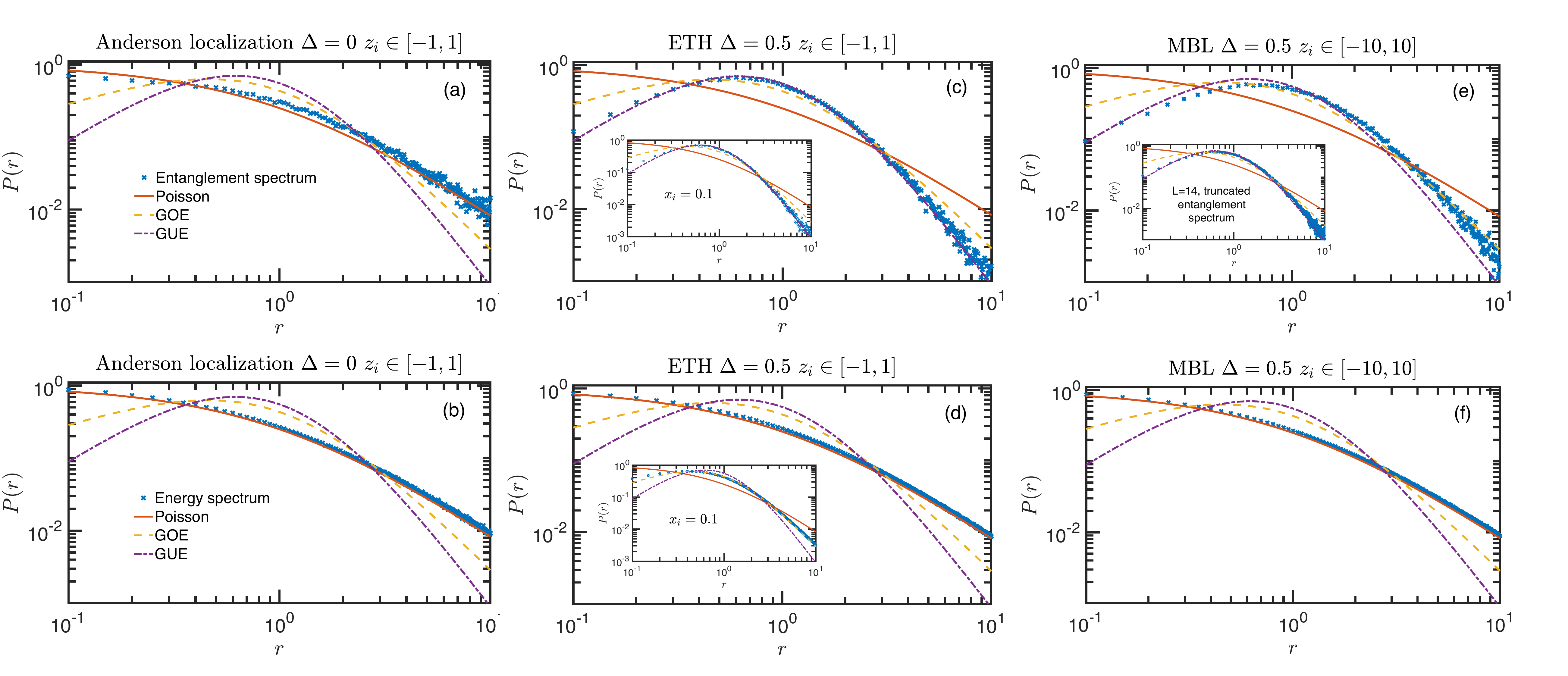}}
\caption{(Color online) Comparison between ESS and energy level
  spacing statistics after a quantum quench at $t_0=1000$ starting
  from a random product state in systems that are Anderson localized
  (a-b), nonintegrable and featuring ETH (c-d), featuring MBL
  (e-f). ESS follows three different distributions, namely Poisson
  (a), WD (c), and a non-universal one (e), thus perfectly classifying
  the three different dynamical phases. On the other hand, the
  distribution of the energy level spacings is always Poisson in all
  three cases. It becomes WD in the nonintegrable, ETH case shown in
  inset of panel (d) only if total magnetization $S_z$ conservation is
  broken by a field in the $x$ direction. In the MBL case, the ESS
  approaches WD upon discarding the largest eigenvalues values of the
  spectrum (inset of (e)). All simulations are done with $2000$
  realizations of disorder and $L=12$ unless otherwise specified.}
\label{weakstrong} 
\end{figure*}

We first consider case (i), the XX spin chain ($\Delta=x_i=0$) in the
presence of a random field $z_i\in[-h,h]$. This model can be brought
into the form of free fermions in one dimension and features AL for
every value of $h$. Here, we choose $h=1$. In Fig.~\ref{weakstrong}a,
we show $P(r)$ of the final states after a long time evolution
($t_0J=1000$). The ESS fits the distribution expected for uncorrelated
eigenvalues, $P_{\rm Poisson}(r) = (1+r)^{-2}$, which can be
straightforwardly derived assuming a Poisson distribution of
spacings. In Refs.~\cite{espectrum1, espectrum2} such statistics
corresponds to simple patterns of entanglement that are easily
reversible under the entanglement cooling algorithm. In the quantum
quench scenario, such pattern results in the failure to reach
thermalization. Indeed, the distribution of the spacings in the energy
spectrum is also Poisson (see Fig.~\ref{weakstrong}b), which is a
typical feature of integrable systems~\cite{berrytabor,rmt}. As we can
see, in the integrable case, the ESS and the energy level spacings
convey the same information. Similarly, we find that in the completely
integrable case ($z_i=0$) both ESS and energy spectrum are still
Poisson. However, because of the absence of localization, entanglement
propagates and fulfills volume law like in a thermal
system~\cite{calabrese}, though no thermalization can happen. This
shows that it is the finer structure of entanglement in the ESS that
is able to diagnose dynamical phases, instead of just the amount of
entanglement.
  
  %%%%%%%%%%%%%%%%%%%%%%%%%%%%%%%%%%%%%%%%%%%%%%%%%%%%%%%%%%%%%%%%%%%%%%%
%%%%%%%%%%%%%%%%%%%%%%%%%%%%%%%%%%%%%%%%%%%%%%%%%%%%%%%%%%%%%%%%%%%%%%%

When the interaction $\Delta$ is switched on, the system can be made
nonintegrable by introducing a random field
$z_i$~\cite{xxz-affleck}. Although nonintegrable, there is still a
simple conserved quantity in the model, namely, the total
magnetization $S_z$ in the $z$ direction. If the disorder is weak (we
choose $h=1$) we are in case (ii): The model obeys ETH and
thermalizes. At this point we are confronted with a shortcoming of the
energy level statistics. For a nonintegrable system, the distribution
of energy level spacings is expected to follow a WD distribution and
very accurate surmises exist in this case~\cite{atas13}: $P_{\rm
  WD}(r) =Z^{-1} ( r + r^2 )^\beta ( 1 + r + r^2 )^{-1-3\beta/2}$,
where $Z={8}/{27}$ for the Gaussian Orthogonal Ensemble (GOE) with
$\beta=1$, and $Z={4\pi}/{81\sqrt{3}}$ for the GUE with
$\beta=2$. However, to find such a result one needs to diagonalize the
Hamiltonian only in the subspace of fixed total
magnetization~\cite{kudo}. If one does not know what the conserved
quantities are -- and this is a generic case -- and diagonalizes the
Hamiltonian in the full Hilbert space, one would find again Poisson
statistics, see Fig.~\ref{weakstrong}d. However, if one breaks the
$S_z$ conservation by a small uniform field in the $x$ direction, one
does find the WD distribution, see inset of
Fig.~\ref{weakstrong}d. Thus, for nonintegrable systems, one is
required to know all conserved quantities in order to check the ETH
through the energy level statistics. The presence of just one (local)
constant of motion makes the system behave as integrable (Poisson
statistics) from the viewpoint of the energy gaps if we consider the
full spectrum, even though the system indeed thermalizes, while
breaking all conservation laws results in WD, see Table~\ref{table1}.

In contrast, we find that the ESS is more robust and captures that
thermalization should not be impaired by the fact that there is one
conserved quantity. We find that the ESS data agrees well with a WD
distribution with $\beta=2$, see Fig.~\ref{weakstrong}c. Breaking the
last constant of motion by introducing a small constant field
$x_i=0.1$ in the $x$ direction results in the same distribution (see
inset). Therefore, it is clear that ESS already gives us an advantage
in comparison to the energy level statistics, as it can discriminate
between integrable and nonintegrable models without requiring the
knowledge of the local conserved quantities.

Finally, keeping fixed $\Delta=0.5$ and increasing the range of $z_i$
we enter in the MBL case (iii). In spite of the system being still
nonintegrable, the energy eigenstates stay very localized breaking
ergodicity and hence thermalization. Moreover, the eigenstates are
weakly entangled (they obey an area law~\cite{arealaw,huse-mbl}, which
for a one-dimensional chain implies an entanglement entropy nearly
independent of the system size). Thus the mechanism behind ETH breaks
down and the system does not thermalize, at least within reasonable
time scales, that is, nonexponential in system size. At such time
scales, the system shows some features of the integrable systems, as
there is an extensive number of quasilocal conserved
quantities~\cite{abanin1, abanin2, huse-mbl,Louk,chandran}. This is
also reflected in the distribution of the energy level spacings. We
computed that distribution and show it in Fig.~\ref{weakstrong}f,
which reveals a Poisson statistics, just like for an integrable system
(or AL, that is, integrable).

\begin{figure}[t]
\centering
\includegraphics[width=.45\textwidth]{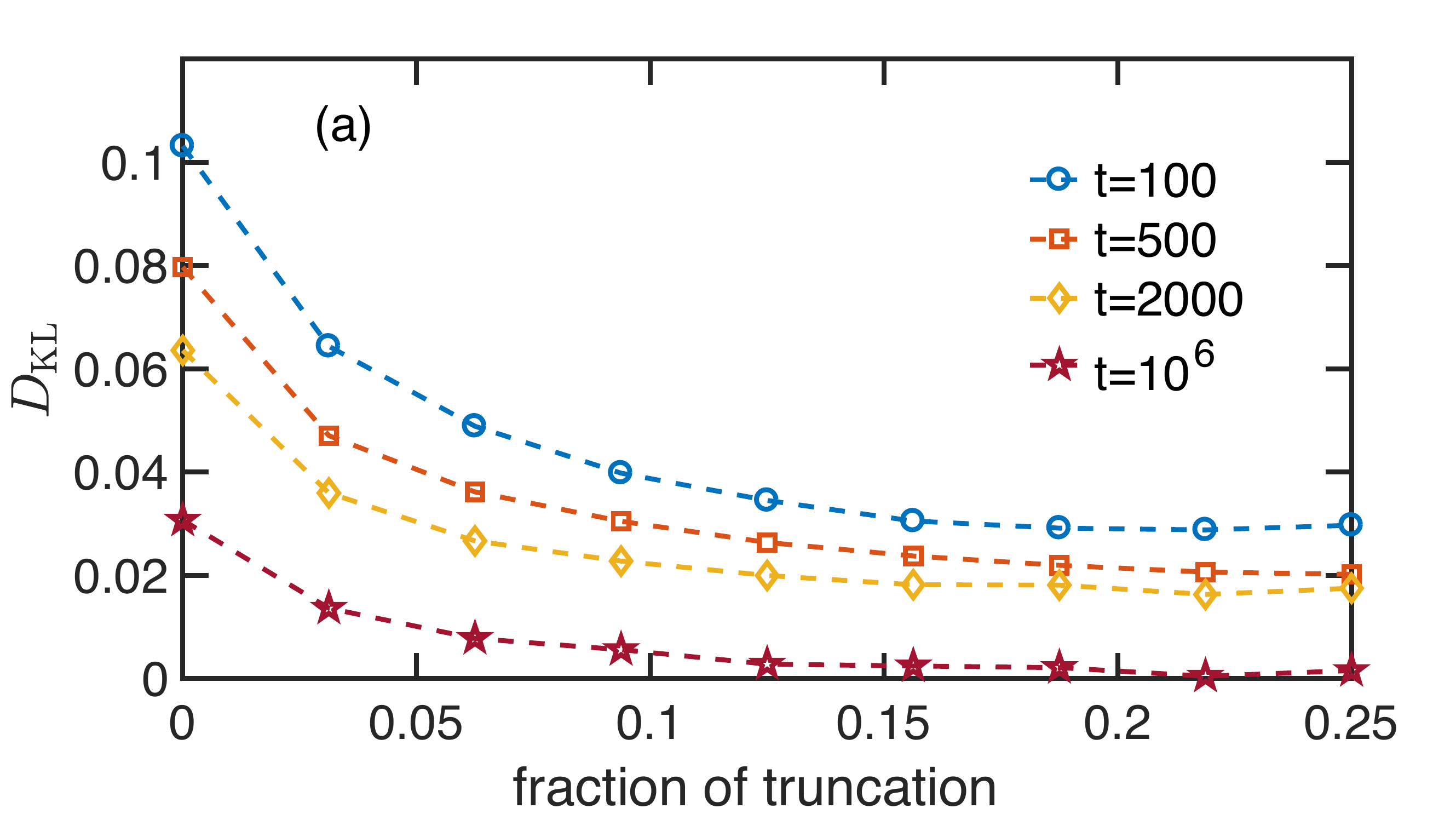} \\
\includegraphics[width=.45\textwidth]{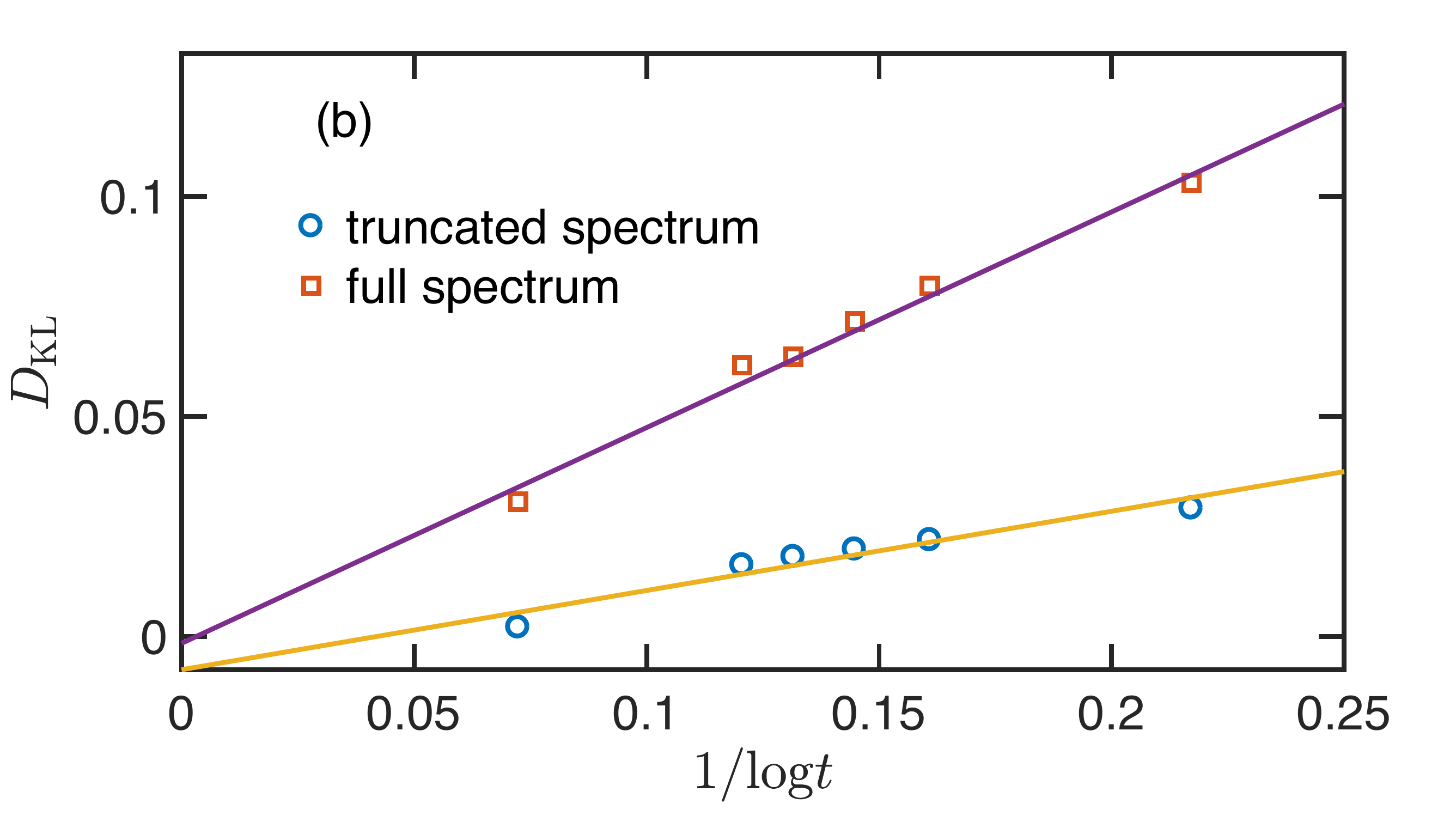} \\
  \caption{(Color online) (a) The KL divergence $D_{\rm KL}$ as
    function of the fraction of truncation of the full spectrum for
    different total evolution times ($L=14$ and $z_i\in[-8,8]$). The
    data are averaged over 100 realizations of disorder and 2000
    realizations of the initial product state, evolved for times
    $t=100, 500, 1000$, and $10^6$. %The values of $D_{\rm KL}$ are reduced as the time increases.
    (b) scaling of $D_{\rm KL}$ with $1/\log(t)$ for the full spectrum
    and for the truncated spectrum at fraction 0.1875, consistent with
    the KL divergence vanishing at long times and the ESS asymtoptically
    reaching the WD distribution.}
  \label{KL}
\end{figure}

Let us now analyze the ESS for MBL. We shall find that MBL can be
distinguished from both AL/integrable systems and ETH. The analysis
that we present below shows that the ESS for MBL approaches
asymptotically a WD distribution at rather long time scales, which we
quantify below. The ESS is shown in Fig.~\ref{weakstrong}e, and show
the following features. At the given time scale ($t_0J=1000$), the ESS
appears to deviate from WD statistics (as well as from Poisson
statistics); the deviation is reduced if one considers a fraction of
the full spectrum, retaining lowest eigenvalues values of the spectrum
and discarding the largest ones (see inset). In order to quantify the
approaching of the entanglement spectrum to WD (GUE) distribution upon
truncation, we consider the statistical distance between two
probability distributions given by the Kullback-Leibler (KL)
divergence: $D_{\rm KL}(p\| q)=\sum_i p_i\log ({p_i}/{q_i})$. In
Fig.~\ref{KL}a, we show the KL divergence between $P(r)$ of MBL and
the WD distribution as function of the fraction of the cutoff. As more
of the largest eigenvalues values are discarded, we get closer to
universal statistics. Moreover, we find that, as function of evolution
time, all the $D_{KL}$ decreases as $1/\log (t)$ (see Fig.~\ref{KL}b),
and thus the ESS of MBL asymptotically approaches a WD (GUE)
distribution. (We remark that the $D_{KL}$ divergence between $P(r)$
and the WD distribution in the ETH regime goes to zero at a time scale
of order $1/J$.) Indeed, in the infinite time limit, time-evolved
states in the MBL regime also have to equilibrate, as the time
fluctuations of typical observables go to zero, though the scaling
with both time and system size are different in MBL from
ETH~\cite{jun}.

%%%%%%%%%%%%%%%%%%%%%%%%%%%%%%%%%%%%%%%%%%%%%%%%%%%%%%%%%%%%%%%%%%%%%%%
%%%%%%%%%%%%%%%%%%%%%%%%%%%%%%%%%%%%%%%%%%%%%%%%%%%%%%%%%%%%%%%%%%%%%%%

We interpret the slow approach to universal WD (GUE) statistics of the
ESS of a state following unitary evolution with a Hamiltonian in the
MBL regime as follows. At reasonable time scales, the system has
approximately local conserved integrals of motion, and may look like
an integrable one. However, unlike AL, the MBL Hamiltonian remains
interacting even in the basis of conserved quantities. Eventually, for
long time scales, information propagates along the full
chain~\cite{friesdorf}, and the interaction between far away
quasilocal conserved quantities is revealed by the slow $1/\log(t)$
approach to the universal WD distribution. The ESS detects the
presence of interaction {\em already} at short time scales, because
the deviations from the universal distribution are small and
decreasing in time. None of these aspects can be captured by the study
of the energy level spacings. We remark that this feature of the ESS
is a truly dynamical one, and depends on the fact that the system is
away from equilibrium. If one truncates the entanglement spectrum of a
high energy \textit{eigenstate} of MBL, the spectrum stays
nonuniversal~\cite{2comp,Nandkishore,powerlaw}.

{\em Complexity of Entanglement.---} The different statistics in the
ESS correspond to different complexity of the entanglement generated
by the time evolution. In Refs.~\cite{espectrum1, espectrum2}, it was
shown that the entanglement generated by a quantum circuit can be
undone by an entanglement cooling algorithm when the ESS shows
(semi-)Poisson statistics. On the other hand, if one uses a quantum
circuit obtained by a universal set of gates, the ESS displays WD
statistics and the simple algorithm for disentangling fails, so the
ESS is complex.

\begin{figure}[t]
\centering
\includegraphics[width=.42\textwidth,clip=true]{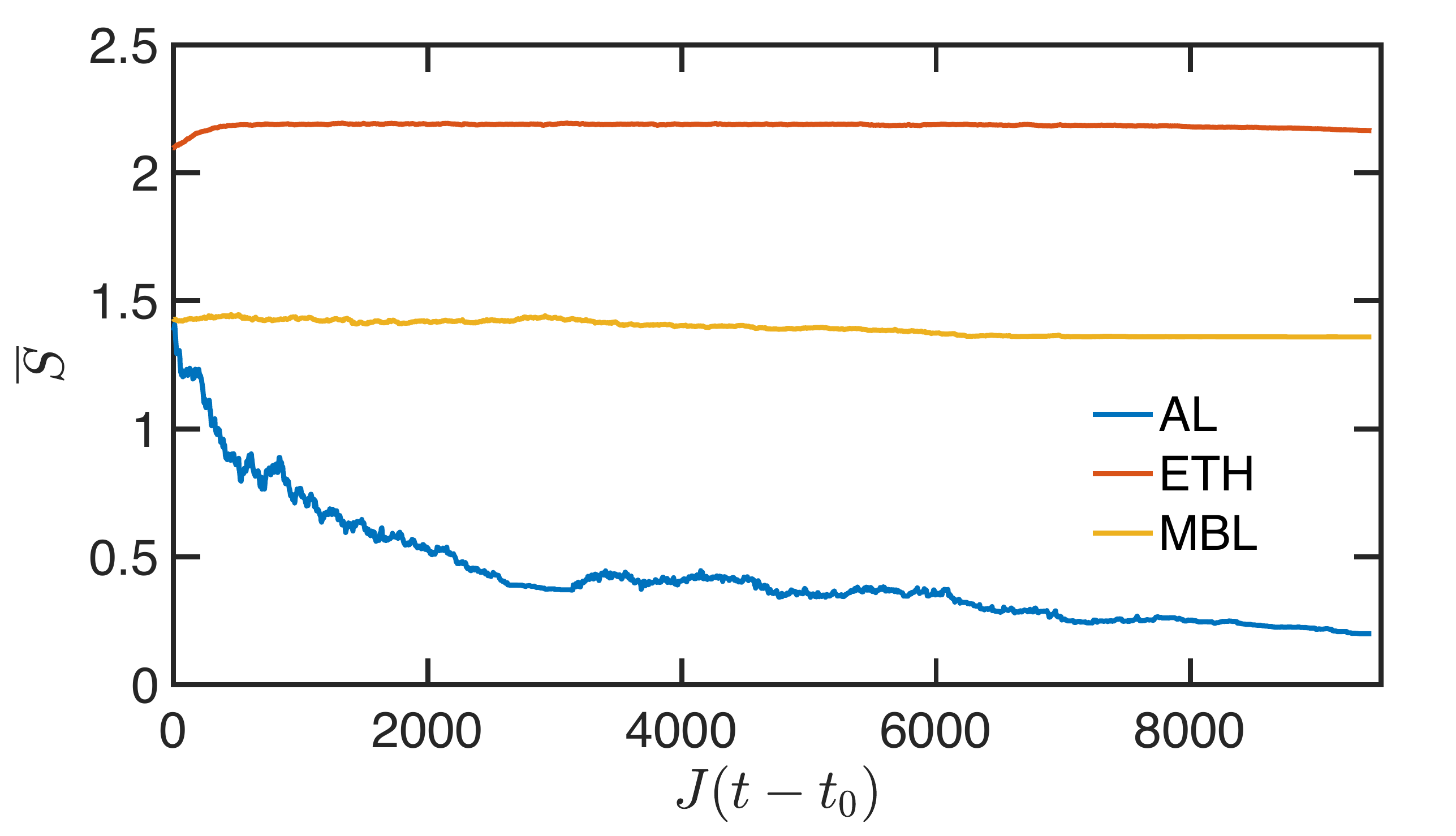}
\caption{(Color online) Attempt of disentangling using the
  entanglement cooling algorithm starting from the states at
  $t_0=1000$. $\bar{S}$ is the von Neumann entropy averaged over all
  possible bipartitions of the system with $L=12$.}
\label{cooling} 
\end{figure}

How does the disentangling algorithm perform in the case of
Hamiltonian evolution? We start from the final state obtained after a
quantum quench for running time $t_0=1000$, like in the previous
analysis for ESS. Notice that a similar amount of entanglement
(averaged over all possible contiguous bipartitions of the system) is
reached in both the MBL and the AL case (see Fig.~\ref{cooling}),
while the average entanglement is much higher for the ETH case. The
disentangling (cooling) algorithm works as follows. We pick randomly a
one-or two-body term from the model Eq.~(\ref{model}), and evolve the
state for a time $\delta t=\pi/10$. Then we accept such an attempt
with probability ${\rm min}\{1, {\rm exp}(-\beta \Delta \bar{S})\}$,
where $\Delta \bar{S}$ is the change of the amount of von Neumann
entropy averaged over all possible bipartitions of the system, and
$\beta^{-1}$ is a fictitious temperature that is gradually reduced to
zero.

%%%%%%%%%%%%%%%%%%%%%%%%%%%%%%%%%%%%%%%%%%%%%%%%%%%%%%%%%%%%%%%%%%%%%%%
%%%%%%%%%%%%%%%%%%%%%%%%%%%%%%%%%%%%%%%%%%%%%%%%%%%%%%%%%%%%%%%%%%%%%%%

Let us look first at the cooling in the disordered XX model, which at
time $t_0=1000$ after the quench features Poisson statistics for the
ESS -- what we would call a non-complex entanglement pattern. The
performance of the cooling algorithm is shown in the blue curve in
Fig.~\ref{cooling}. As the data show, the state can be disentangled
almost completely by this kind of entanglement cooling algorithm. It
is a remarkable fact that entanglement can be undone after Hamiltonian
evolution even without knowledge of the precise Hamiltonian.

What happens for ETH and MBL? Figure~\ref{cooling} shows that the
entanglement entropy reached at $t_0=1000$ using both the MBL and ETH
Hamiltonians cannot be undone by the cooling algorithm, even though
the value of the entanglement entropy is smaller in the case of
MBL. States generated from evolutions using MBL or ETH Hamiltonians
cannot be disentangled, and in both cases, the ESS shows some degree
of universality (both reach a WD distribution, albeit at rather
different time scales). We conclude that what determines how easy or
hard it is to disentangle a state is not the level of entanglement, as
measured by the entanglement entropy, but instead that information is
contained in the ESS, like in the case for states generated by quantum
circuits.

This work is supported by DOE Grant DE-FG02-06ER46316 (Z.-C.Y. and
C.C.).

%%%%%%%%%%%%%%%%%%%%%%%%%%%%%%%%%%%%%%%%%%%%%%%%%%%%%%%%%%%%%%%%%%%%%%%

%%%%%%%%%%%%%%%%%%%%%%%%%%%%%%%%%%%%%%%%%%%%%%%%%%%
%%%%%%%%%%%%%%%%%%%%%%%%%%%%%%%%%%%%%%%%%%%%%%%%%%%

\end{document}